\documentclass[twocolumn,11pt]{article}
\usepackage{times}

\usepackage{amssymb, amsmath}
\usepackage{amsthm}
\usepackage{amsfonts}
\usepackage{latexsym}
\usepackage{graphicx}
\usepackage{epsf}
\usepackage{listings}
\newtheorem{definition}{Definition}

\newtheorem{lemma}[definition]{Lemma}
\newtheorem{theorem}[definition]{Theorem}

\newtheorem{example}[definition]{Example}
\newcommand{\N}{\mathbb{N}}
\newcommand{\Nplus}{\mathbb{N}^{+}}
\begin{document}
\global\def\refname{{\normalsize \it References:}}
\baselineskip 12.5pt
%
%
% TITLE, AUTHOR, ABSTRACT, KEYWORDS
%
\title{\LARGE \bf Some Problems and Algorithms Related to the Weight Order Relation on the \emph{n}-dimensional Boolean Cube}

\date{}

\author{\hspace*{-10pt}
\begin{minipage}[t]{2.7in} \normalsize \baselineskip 12.5pt
\centerline{VALENTIN BAKOEV}
\centerline{``St. Cyril and St. Methodius'' University of V. Tarnovo}
\centerline{Faculty of Mathematics and Informatics}
\centerline{2 Theodosii Tarnovski Str., 5003 Veliko Tarnovo}
\centerline{BULGARIA}
\centerline{v.bakoev@ts.uni-vt.bg}
\end{minipage} \kern 0in
\\ \\ \hspace*{-10pt}
\begin{minipage}[b]{6.9in} \normalsize
\baselineskip 12.5pt {\it Abstract:}
	The problem ``Given a Boolean function $f$ of $n$ variables by its truth table vector. Find (if exists) a vector $\alpha \in \{0,1\}^n$ of maximal (or minimal) weight, such that $f(\alpha)= 1$.'' is considered here. It is closely related to the problem of fast computing the algebraic degree of Boolean functions. It is an important cryptographic parameter used in the design of S-boxes in modern block ciphers, pseudo-random numbers generators in stream ciphers, at Reed-Muller codes, etc. To find effective solutions to this problem we explore the orders of the vectors of the $n$-dimensional Boolean cube $\{0,1\}^n$ in accordance with their weights. The notion of ``$k$-th layer'' of $\{0,1\}^n$ is involved in the definition and examination of the ``weight order'' relation. It is compared with the known relation ``precedes''. Several enumeration problems concerning these relations are solved and the corresponding comments were added to 3 sequences in the On-line Encyclopedia of Integer Sequences (OEIS). One special order (among the numerous weight orders) is defined and examined in detail. The lexicographic order is a second criterion for an ordinance of the vectors of equal weights. So a total order called  Weight-Lexicographic Order (WLO) is obtained. Two algorithms for generating the WLO sequence and two algorithms for generating the characteristic vectors of the layers are proposed. Their results were used in creating 2 new sequences: A294648 and A305860 in the OEIS. Two algorithms for solving the problem considered are developed---the first one works in a byte-wise manner and uses the WLO sequence, and the second one works in a bitwise manner and uses the characteristic vector as masks. The experimental results from numerous tests confirm the efficiency of these algorithms. Some other applications of the obtained algorithms are also discussed---for example, when representing, generating and ranking other combinatorial objects.
\\ [4mm] {\it Key--Words:}
% The key-words follow.
Boolean cube, binary vector, serial number, lexicographic order, weight order, characteristic vector, layer, maximum chains enumerating, weight-lexicographic order, generating algorithm, search algorithm
\end{minipage}
\vspace{-10pt}}

% \emph{Mathematics Subject Classification}:
% Primary 68R05; Secondary	06A07, 05A15, 05A18.

\maketitle

\thispagestyle{empty} \pagestyle{empty}
% numbers of pages are supplemented by the editor
%
% THE BEGINNING OF THE TEXT
%
\section{Introduction}
\label{Intro} \vspace{-4pt}
	The algebraic degree of Boolean function (or vectorial Boolean function, called S-box) is an important cryptographic parameter. It is used in the design of S-boxes for modern block ciphers, pseudo-random numbers generators in stream ciphers, at the Reed-Muller codes, etc. \cite{CC_BFCECC, CC_VBFC, ACANT}. The algorithms that compute this parameter (as well as the other cryptographic parameters) must be very fast since when generating such examples, this parameter is computed for each of them. As faster is the algorithm, more examples can be generated and a better choice among them to be done. 
	
	The problem we consider here is closely related to the problem of computing the algebraic degree of Boolean function. It is: ``Given a Boolean function $f$ of $n$ variables by its Truth Table vector, denoted by $TT(f)$. Find (if exists) a vector $\alpha \in \{0,1\}^n$ of maximal (or minimal) weight, such that $f(\alpha)= 1$.''. For brevity, we call this problem \textit{VectorOfMaxWeight}. The simplest way to solve it is to perform an \emph{exhaustive} (\emph{linear}) \emph{search}: for each vector $\beta \in \{0,1\}^n$ it checks whether $f(\beta)=1$ and selects the vector of maximal (resp. minimal) weight. Since the values (coordinates) in $TT(f)$ correspond to the lexicographic order of the vectors of $\{0,1\}^n$, the algorithm cannot stop before to check each coordinate of $f$. So, it performs $\Theta (2^n)$ checks. However, if the values of $TT(f)$ are checked in accordance with the vectors' weights, from the highest to the lowest weight, the search will finish after finding the first vector $\beta \in \{0,1\}^n$, such that $f(\beta)=1$. Once the desired weight order of the vectors has been obtained, this approach needs $O(2^n)$ checks. This order can be obtained by an algorithm that: (1) computes the vectors' weights and (2) sorts the vectors in accordance with their weights. So it needs at least $\Theta (n.2^n)$ operations. Instead of this simple solution, here we investigate the properties of the weight order relation defined on $\{0,1\}^n$. These properties are applied to solve some enumeration problems and have useful generalizations and applications that are the basis for creating more efficient algorithms. Another approach for computing the algebraic degree of Boolean function $f$, that investigates the support of $f$, derives and uses its algebraic properties, is proposed in \cite{CGV}. 
	
  This paper represents the comprehensive study of the subject under discussion and the results obtained so far (the first of them were reported in \cite {VB2018}). It is organized as follows. The necessary basic notions concerning the Boolean cube and their properties are given in Section \ref{Basic_Notions}. In Section \ref{WO_Relation}, the  relation ``precedes by weight'' is defined, examined and compared with the known relation ``precedes''. Some enumeration problems concerning both relations are solved and the corresponding notes were added to the sequences A051459, A001142 and A000142 in the OEIS \cite{SLO}. In Section \ref{WLO_Relation} one special order called a Weight-Lexicographic Order (WLO) is introduced and explored. Two algorithms for generating the WLO sequence $l_n$ are proposed. A third algorithm, that uses the sequence $l_n$ in solving the problem VectorOfMaxWeight, is also proposed and discussed. The results in this section were used in creating the sequence A294648 in the OEIS \cite{SLO}. In Section \ref{Gen_CV_WO_Rel}, the characteristic vectors of the layers and their generating are considered. A bitwise version of an algorithm for solving the problem VectorOfMaxWeight is discussed, where these vectors are used as masks. The sequence A305860 in OEIS \cite{SLO} was created by using the results from this section. Section \ref{Exp_Res} shows the experimental results from numerous tests conducted for comparison of the efficiency of algorithms discussed. The results show convincingly the superiority of the algorithms based on WLO. The last section contains comments and explanations about the algorithms based on WLO and their applications in computing the algebraic degree of Boolean functions. Some applications of the  obtained algorithms in generating other combinatorial objects, their representations, ranking/unranking are also discussed.
\section{Basic notions, properties}
\label{Basic_Notions}
\vspace{-4pt}
	The necessary basic concepts about the Boolean cube and their properties are represented following \cite{VB2014}. Let $\N$ be the set of natural numbers,  and $\N^{+}=\N \backslash \{0\}$ be the set of positive natural numbers.

	Usually, the \textit{$n$-dimensional Boolean cube} is defined as $\{0,1\}^n= \{(x_1,x_2, \dots, x_n)| x_i\in \{0,1\}, i=1,2,\dots, n \}$, i.e., the set of all $n$-dimensional binary vectors. So their number is $|\{0,1\}^n| = |\{0,1\}|^n= 2^n$. However, the following alternative, inductive and constructive definition is more useful further.

\begin{definition}
\label{D10}
\normalfont{
	1) The set $\{0,1\}= \{(0),(1)\}$ is called \textit{one-dimensional 
Boolean cube} and its elements $(0)$ and $(1)$ are called \textit{one-dimensional binary vectors}.
\\
	2) Let $\{0,1\}^{n-1}=\{ \alpha_0, \alpha_1, \dots$, $\alpha_{2^{n-1}-1} \}$ be the \textit{$(n-1)$-dimensional Boolean cube} and $\alpha_0, \alpha_1, \dots,$ $\alpha_{2^{n-1}-1}$ be its \textit{$(n-1)$-dimensional binary vectors}. 
\\
	3) The \textit{$n$-dimensional Boolean cube} $\{0,1\}^n$ is built by taking the vectors of $\{0,1\}^{n-1}$ twice: firstly, each vector of $\{0,1\}^{n-1}$ is prefixed by zero, and thereafter each vector of $\{0,1\}^{n-1}$ is prefixed by one, i.e.,
\begin{align*}
\{0,1\}^n &=&\{(0, \alpha_0),(0,\alpha_1), \dots ,(0,\alpha_{2^{n-1}-1}),\\
          & &(1, \alpha_0),(1,\alpha_1), \dots ,(1,\alpha_{2^{n-1}-1}) \}.	
\end{align*}
}
\end{definition}

\begin{figure}[ht]
    \centering
       \scalebox{0.8}{\includegraphics{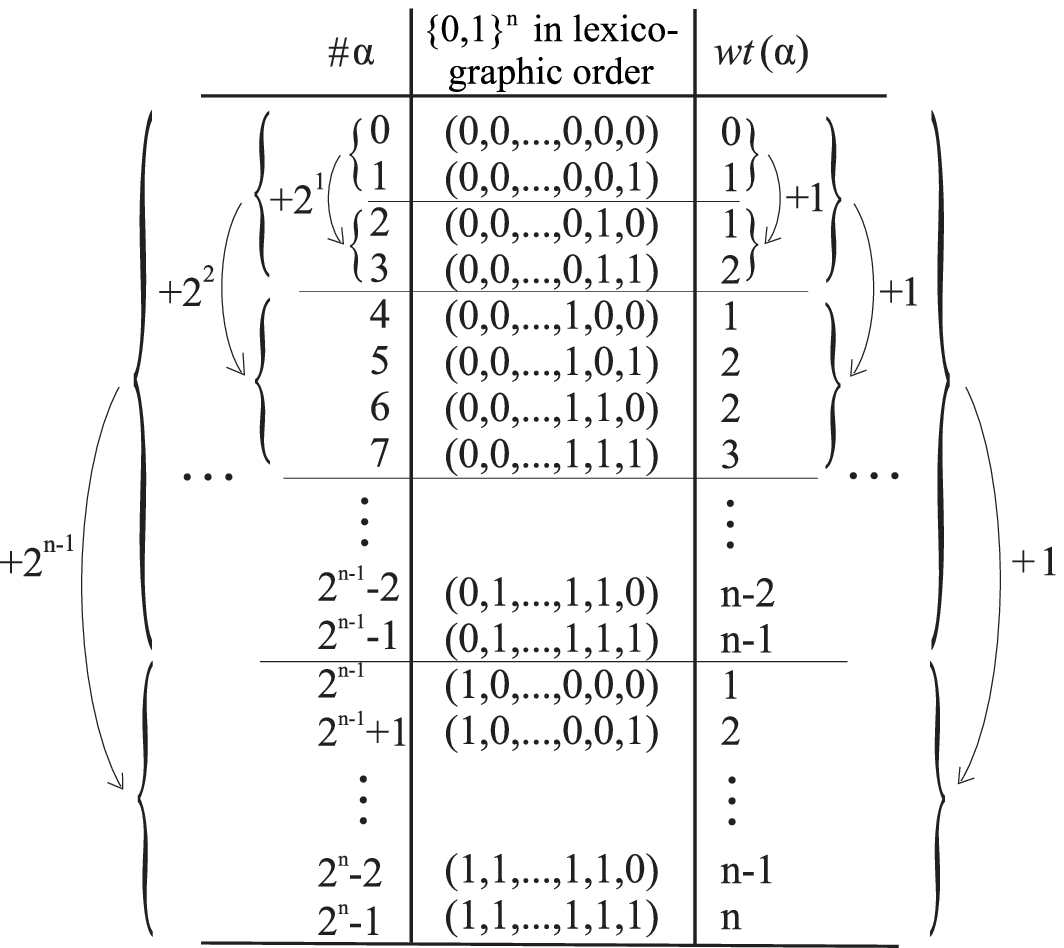}}
    \caption{Illustration of the statement of Theorem \ref{T10}}
    \label{fig:Lex_N_Wt-Relation}    
\end{figure}

	Let $\alpha=(a_1, a_2, \dots, a_n)\in \{0,1\}^n$ be an arbitrary vector. 
The natural number $\#\alpha=\sum_{i=1}^n a_i.2^{n-i}$ is called a  \textit{serial number} of the vector $\alpha$. So $\#\alpha$ is the natural number whose $n$-digit binary representation is $a_1 a_2 \dots a_n$. A \textit{weight} (or \textit{Hamming weight}) of $\alpha$ is the natural number $wt(\alpha)$, equal to the number of non-zero coordinates of $\alpha$, i.e., $wt(\alpha)= \sum_{i=1}^n a_i$. These and some of the following notions are illustrated in Figure \ref{fig:Lex_N_Wt-Relation} and Example \ref{Ex10}.
\begin{definition}
\label{D60}
\normalfont{
	For arbitrary vectors $\alpha=(a_1, a_2,\dots,$ $a_n)$ and $\beta=(b_1, b_2,$$\dots, b_n)$ $\in \{0,1\}^n$, the relation \textit{lexicographic precedence} $R_{\leq} \subseteq \{0,1\}^n \times \{0,1\}^n$ is defined as follows: $(\alpha, \beta) \in R_{\leq}$, if $\alpha= \beta$ or $\exists\, i, 1\leq i\leq n$, such that $a_i<b_i$, and $a_j=b_j$ for $j< i$. When $(\alpha, \beta) \in R_{\leq}$ we say that $\alpha$ \textit{lexicographically precedes} $\beta$ and write $\alpha \leq \beta$.
}
\end{definition}

	The relation $R_{\leq}$ is reflexive, antisymmetric and transitive. Furthermore, each pair of vectors $\alpha, \beta \in \{0,1\}^n$ are \textit{comparable} with respect to $R_{\leq}$, i.e., either $\alpha\leq \beta$, or $\beta \leq \alpha$ holds. So $R_{\leq}$ is a \textit{total order} in $\{0,1\}^n$. This means that its vectors can be \textit{ordered} (or \textit{sorted}) \textit{lexicographically} in a unique way in the sequence $\alpha_0, \alpha_1,\dots, \alpha_k$, $\dots$, $\alpha_{2^n-1}$, such that $\alpha_l \leq \alpha_k$, $\forall\, l<k$, and $\alpha_k \leq \alpha_r$, $\forall\, k<r$, and for any $k=0, 1,\dots, 2^n-1$. 
\begin{theorem}
\label{T10}
	If the vectors of $\{0,1\}^n$  are obtained in accordance with Definition \ref{D10}, then:

1) They are in lexicographic order.

2) The serial numbers of the vectors form the sequence of natural numbers: $0, 1, \dots,$ $2^n-1$. So $\alpha \leq \beta$ if and only if $\#\alpha \leq \#\beta$.

3) The weights of the vectors in the second half of $\{0,1\}^n$ are obtained 
by adding 1 to the weights of corresponding vectors from the first half of 
the cube.
\end{theorem}

	Following Definition \ref{D10}, the proof of the theorem by induction on $n$ is easy and that is why it is omitted. The theorem states the bijection between the vectors in lexicographic order and their serial numbers, i.e., the vectors of $\{0,1\}^n$ are in lexicographic order if and only if the sequence of their serial numbers is $0, 1, \dots, 2^n-1$. It also shows the relation between the vectors in lexicographic order and their weights. Its assertions are illustrated in Figure \ref{fig:Lex_N_Wt-Relation}. The right column is the sequence A000120 in the OEIS \cite{SLO}, titled ``1's-counting sequence: number of 1's in binary expansion of n (or the binary weight of n).''. This column shows that the sequence of weights for the lower half of each subcube is obtained by the addition of the number 1 to the corresponding terms of the sequence for the upper half of the same subcube, following the third (inductive) step of Definition \ref{D10}. These relations can be used for efficient computing of the vectors' weights of $\{0,1\}^n$, as in \cite{IBVB}. Analogously, the left column shows that the sequence of serial numbers in the lower half of each subcube is obtained from the corresponding sequence of the upper half of the same subcube by consecutive addition of the number $2^1, 2^2,\dots, 2^{n-1}$, following the third step of Definition \ref{D10} again.

\begin{figure}[ht]
    \centering
        \scalebox{0.7}{\includegraphics{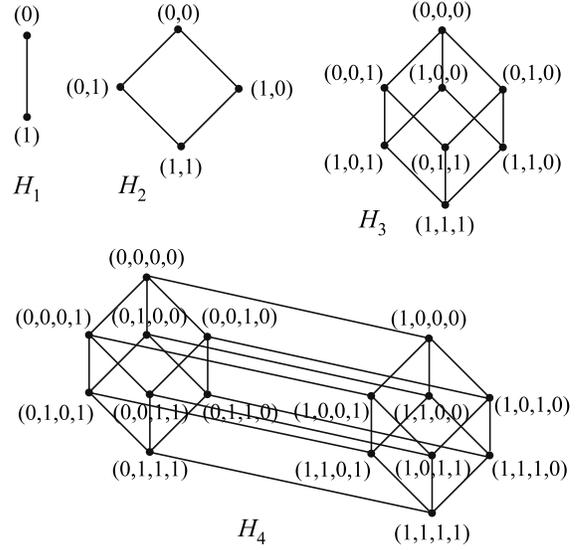}}
    \caption{The graphs $H_1, \dots, H_4$}
    \label{fig:B_Cube-Graph}    
\end{figure}

	Let $\alpha=(a_1, a_2, \dots, a_n)$ and $\beta=(b_1, b_2, \dots,$ $b_n)$ be 
arbitrary vectors of $\{0,1\}^n$. A \textit{Hamming distance} between $\alpha$ and $\beta$ is the natural number $d(\alpha, \beta)$ equal to the number of coordinates in which $\alpha$ and $\beta$ differ. If $d(\alpha, \beta)= 1$, then $\alpha$ and $\beta$ are called \textit{adjacent}, or more precisely \textit{adjacent in $i$-th coordinate} if they differ in this coordinate only. If $d(\alpha, \beta)= n$, the vectors $\alpha$ and $\beta$ are called \textit{opposite} to each other. The \textit{graph} of the $n$-dimensional boolean cube is defined as $H_n=(V_n, E_n)$, where $V_n=\{0,1\}^n$ and $E_n=\{\{\alpha, \beta\}|\, \alpha, \beta\in\{0,1\}^n : d(\alpha , \beta)=1\}$, i.e., the vectors of the cube are vertices of $H_n$ and each pair adjacent vectors are connected by an edge. The graphs $H_1,\dots, H_4$ are shown in Figure \ref{fig:B_Cube-Graph}. 
\begin{definition}
\label{D90}
\normalfont{
	The \textit{precedence} relation is denoted by $R_{\preceq}$ and it is defined as follows: for arbitrary vectors $\alpha=(a_1, a_2,\dots, a_n)$, $\beta=(b_1, b_2,\dots, b_n)$ $\in \{0,1\}^n$, $(\alpha, \beta) \in R_{\preceq}$ if $a_i\leq b_i, \forall\, i=1,2,\dots, n$. When $(\alpha, \beta) \in R_{\preceq}$ we say that $\alpha$ \textit{precedes} $\beta$ and write $\alpha \preceq\beta$. When $\alpha \preceq\beta$ or $\beta \preceq \alpha$ the vectors $\alpha$ and $\beta$ are called \textit{comparable}, and otherwise---\textit{incomparable}.
}
\end{definition}

	$R_{\preceq}$ is reflexive, antisymmetric and transitive, and so it is a \textit{partial order} in $\{0,1\}^n$. Thus $\{0,1\}^n$ is a \textit{partially ordered set} (POSet) with respect to $R_{\preceq}$. It is denoted by $(\{0,1\}^n, R_{\preceq})$ or simply by $(\{0,1\}^n, \preceq)$. $R_{\preceq}$ is not a total order because not all pairs $\alpha$, $\beta \in \{0,1\}^n$ are comparable---for example, all vectors of equal weights are incomparable. 

	The vector $\alpha\in \{0,1\}^n$ is called a \textit{minimal element} of the POSet $(\{0,1\}^n,$ $R_{\preceq})$, if $\alpha \preceq \beta$, for any $\beta \in \{0,1\}^n$. Analogously, the vector $\delta\in \{0,1\}^n$ is called a \textit{maximal element} of $(\{0,1\}^n,$ $R_{\preceq})$, if $\gamma \preceq\delta$, for any $\gamma\in \{0,1\}^n$. So, the \textit{zero vector} $\vec{0}_n= (0,0,\dots, 0)$ of $n$-coordinates and the \textit{all-ones vector} $\vec{1}_n=(1,1,\dots, 1)$ of $n$-coordinates are the minimal and the maximal elements of the POSet $(\{0,1\}^n, \preceq)$, correspondingly. If any pair of vectors of the subset $C\subset \{0,1\}^n$ are comparable, they can be ordered in a unique way in a \textit{chain}, for example $\alpha_{i_1}, \alpha_{i_2}, \dots,  \alpha_{i_k}, \dots , \alpha_{i_m}$, such that $\alpha_{i_l} \preceq \alpha_{i_k}$, for $l<k$, and $\alpha_{i_k} \preceq \alpha_{i_r}$, for $k<r$, and for $k=1, 2,\dots, m$. A chain that is not a proper subset of any other chain is a \textit{maximal chain}. For example, $(0,0,0)$, $(0,1,0)$, $(1,1,0)$, $(1,1,1)$ is a maximal chain in $\{0,1\}^3$, whereas $(0,1,0,0)$, $(0,1,0,1)$, $(1,1,0,1)$, $(1,1,1,1)$ is not a maximal chain in $\{0,1\}^4$---see Figure \ref{fig:B_Cube-Graph}. The maximal chain should contain the minimal and  maximal element of the corresponding POSet. Each chain of greatest possible size is called a \textit{maximum} (or \textit{longest}) chain.
\begin{definition}
\label{D130}
\normalfont{
	Let $U=\{x_1,x_2,\dots, x_n\}$ be a given set, $n\in\Nplus$, and $X\subseteq U$. The vector $\alpha= (a_1,a_2,$ $\dots,a_n)\in \{0,1\}^n$, defined as:
\begin{eqnarray*}
a_i=\left\{
\begin{array} {ll}
0, \textrm{\ if\ } x_i\notin X\,,\\
1, \textrm{\ if\ } x_i\in X\,,
\end{array}
\right.
\end{eqnarray*} 
for $i=1, 2, \dots ,n$, is called a \textit{characteristic vector} of the set $X$. 
}
\end{definition}
\begin{example}
\label{Ex10}
\normalfont{
	Let $U=\{a,b,c,d,e,f\}$, $X=\{b, c, e\}$ and $Y=\{c, a, f, d\}$. Since $|U|=6$, $\alpha=(0,1,1,0,1,0)\in \{0,1\}^6$ is the characteristic vector of $X$, and $\beta= (1,0,1,1,0,1)$---the characteristic vector of $Y$. The vectors $\gamma=\vec{0}_6$ and $\delta=\vec{1}_6$ are the characteristic vectors of $\emptyset \subseteq U$ and $U\subseteq U$, correspondingly. Furthermore:
\begin{itemize}
	\item $\#\alpha= 26$,\ $\#\beta= 45$,\ $\#\gamma= 0$,\ $\#\delta= 65$;
	\item $wt(\alpha)= 3,\  wt(\beta)= 4,\ wt(\gamma)= 0,\ wt(\delta)= 6$;
	\item $d(\alpha, \gamma)= 3,\ d(\alpha, \beta)= 5,\ d(\beta, \delta)= 2$, etc.;
	\item $\gamma\leq \alpha \leq \beta\leq \delta$, in accordance with Definition \ref{D60} and Theorem \ref{T10};
	\item $\gamma \preceq \alpha, \gamma \preceq \beta, \alpha \preceq \delta, \beta \preceq \delta$, etc., but $\alpha$ and $\beta$ are incomparable with respect to $R_{\preceq}$.
\end{itemize}
}
\end{example}
\begin{theorem}
\label{T20}
	Let $U$ be an $n$-element set, $n\in\Nplus$, and ${\mathcal{P}}(U)$ be the power set of $U$. Let  $f:{\mathcal{P}}(U) \rightarrow$ $\{0,1\}^n$ be a function defined as follows: $f(X)=\alpha$, where $\alpha \in \{0,1\}^n$ is the characteristic vector of $X$, for any $X\in {\mathcal{P}}(U)$. Then $f$ is a bijection.
\end{theorem}

	The proof of the theorem is omitted because it is trivial. The function $f$ from Theorem \ref{T20} bijectively relates (maps) the bitwise operations on the binary vectors to the operations on the subsets of a given $n$-element set $U$ as follows: $\vee$  (disjunction) and $\cup$ (union); $\wedge$ (conjunction) and $\cap$ (intersection); $\overline{\raisebox{1ex}{\ \ }}$ (negation) and $\overline{\raisebox{1ex}{\ \ }}$ (complement); $\oplus$ (sum modulo 2, XOR) and $\Delta$ (symmetric difference), correspondingly. These properties are generalized in the following theorem \cite{KUZNO, GRIM, GRTJ}.
\begin{theorem}
\label{T30}
	Let $U$ be an $n$-element set, $n\in \N^{+}$. Then the Boolean algebras $({\mathcal{P}}(U)$, $\cup, \cap, \overline{\raisebox{1ex}{\ \ }}, \emptyset, U)$	and $(\{0,1\}^n, \vee, \wedge,\overline{\raisebox{1ex}{\ \ }}, \vec{0}_n, \vec{1}_n)$ are isomorphic.
\end{theorem}

	Furthermore, the bijection $f$ from Theorem \ref{T20} concerns the relations $R_\subseteq$ (defined on a given universal set  $U$, $|U|=n\in \Nplus$) and $R_\preceq$ (defined on $\{0,1\}^n$). For arbitrary $A, B\subseteq U$, having  characteristic vectors $\alpha, \beta \in \{0,1\}^n$, correspondingly, it is easy to prove that $A\subseteq B \Leftrightarrow \alpha \preceq \beta$, i.e.,
$(A, B)\in R_\subseteq$ $\Leftrightarrow (f(A), f(B))\in R_\preceq$.
Thus $f$ is an \textit{isomorphism} between the POSets $({\mathcal{P}}(U), R_\subseteq)$ and $(\{0,1\}^n$, $R_\preceq)$ that preserves the relations and the orders corresponding to them. This property is illustrated in Figure \ref{fig:WLO-B_cube-Subsets} by the graphs of the corresponding relations, for $n=3$. 

	These important \textit{structural properties} are used in \cite{KNUTH, AUH1, ANHW, FRRUSK, DKDS, RND, CLRS, SSKI}, etc., for:
\begin{itemize}
	\item Computer representations of sets by binary vectors or arrays and performance of the basic operations on them. The concepts of characteristic vector and serial number, bijectively related by Theorem \ref{T20}, are used for \textit{ranking/unranking} of the subsets of a given universal set and this is the most natural ranking/unranking function.
	\item Generating all subsets (or the $k$-element subsets, $k$-combinations) of a given $n$-element set in a definite order. 
\end{itemize}

	The following exposition is related to all these properties and applications.
\section{The weight-order relation and enumeration problems related to it}
\label{WO_Relation}
\vspace{-4pt}
\begin{definition}
\label{D40}
\normalfont{
	For an arbitrary $k\in \N, k\leq n$, the set of all $n$-dimensional binary vectors of weight $k$ is called a \textit{$k$-th layer} of the $n$-dimensional Boolean cube. We denote it by $L_{n,k}=\{\alpha|\, \alpha \in \{0,1\}^n : wt(\alpha)=k\}$.
}
\end{definition}

	Figure \ref{fig:B_Cube-Graph} illustrates the notion of layer from Definition \ref{D40}. All vectors in the same horizontal level in the figure form the corresponding layer of the cube. Since $k$ coordinates can be chosen among $n$ coordinates (and filled in with units) in $\binom{n}{k}$ ways, hence $|L_{n,k}|= \binom{n}{k}$, for $k=0, 1, \dots, n$.  The family of all layers $L_n=\{ L_{n,0}, L_{n,1},\dots,L_{n,n}\}$ is a \textit{partition} of the $n$-dimensional Boolean cube into layers and hence:
\[
\left|\bigcup_{i=0}^n L_i \right| = \sum_{k=0}^n |L_i|= \sum_{k=0}^n \binom{n}{k}= 2^n= |\{0, 1\}^n|\,.
\]
	 Moreover, the \textit{sequence of layers} $L_{n,0}, L_{n,1}, \dots, L_{n,n}$ is an order of the vectors of $\{0,1\}^n$ in accordance with their weights. So, if $\alpha, \beta \in \{0,1\}^n$ and $wt(\alpha) < wt(\beta)$, then $\alpha$ precedes $\beta$ in the sequence of layers, and when $wt(\alpha)=wt(\beta)=k$, then $\alpha, \beta \in L_{n,k}$ and there is no precedence between them. More precisely, the corresponding relation $R_{<_{wt}}$ can be defined as follows: for arbitrary $\alpha, \beta \in \{0, 1\}^n$, $(\alpha, \beta) \in R_{<_{wt}}$ if $wt(\alpha) < wt(\beta)$. We want $R_{<_{wt}}$ to be reflexive and we set $(\alpha, \alpha) \in R_{<_{wt}}$. When $(\alpha, \beta) \in R_{<_{wt}}$ we say that ''$\alpha$ \textit{precedes by weight} $\beta$'' and write also $\alpha <_{wt} \beta$. It is easy to verify that $R_{<_{wt}}$ is a partial order in $\{0, 1\}^n$ and we refer to it as a \textit{Weight-Order} (WO) further. 

	The vectors of $L_{n,k}$ can be rearranged in $\binom{n}{k}!$ ways, for $k= 0, 1, \dots, n$. Thus we obtain $\prod_{k=0}^n \binom{n}{k}!$ ways for WO of the vectors of $\{0,1\}^n$. The product values obtained for $n=1,2,3,4, \dots$ are $1, 2, 36, 414720,\dots$, correspondingly. They form the sequence A051459 in the OEIS \cite{SLO}, which is defined by Yuval Dekel (Nov 15 2003) very shortly as ``Number of orderings of the subsets of a set with $n$ elements that are compatible with the subsets' sizes; i.e., if $A$, $B$ are two subsets with $A < = B$ then $Card(A) < = Card(B)$''. This description corresponds to the assertion of Theorem \ref{T20} and to the notion WO, since the vectors in the layer $L_{n,k}$ are characteristic vectors of all $k$-element subsets of an $n$-element set, for $k= 0, 1, \dots, n$. In addition, we note that $\prod_{k=0}^n \binom{n}{k}!$ is the number of \textit{all possible topological orders} (or \textit{sorts}) of the directed acyclic graph defined by the same POSet. The corresponding comments were added to the sequence A051459.
\begin{theorem}
\label{T40}
	The number of maximum chains in the POSet $(\{0,1\}^n, R_{<_{wt}})$ is equal to $\prod_{k=0}^n \binom{n}{k}$\,, for any $n\in \Nplus$.
\begin{proof}
\normalfont{
	We consider the POSet $(\{0,1\}^n, R_{<_{wt}})$, for arbitrary $n\in \Nplus$. A maximum chain cannot contain 2 or more vectors from the same layer because there is no precedence by weight between any two vectors from the same layer. So the length of any maximum chain is equal to the number of layers in $\{0,1\}^n$, which is $n+1$. For each $k= 0,1,\dots, n$, there are $\binom{n}{k}$ ways to choose a vector from $L_{n,k}$ which to participate in a maximum chain. Following the multiplication rule, there are $\prod_{k=0}^n \binom{n}{k}$ maximum chains in this POSet.
}
\end{proof}
\end{theorem}	

The formula $\prod_{k=0}^n \binom{n}{k}$ means the product of binomial coefficients from the $n$-th row of Pascal triangle. Its values obtained for $n= 1,2,3,4,5 \dots$ are $1, 2, 9, 96, 2500 \dots$, correspondingly, and they  form the sequence A001142 in the OEIS \cite{SLO}. The assertion of Theorem \ref{T40} was added in the description of A001142.

Let us consider the connection between the relations $R_{<_{wt}}$ and  $R_{\preceq}$. We note that $\alpha \preceq \beta$ always implies $\alpha <_{wt} \beta$. However, $\alpha <_{wt} \beta$ does not imply $\alpha \preceq \beta$ in the general case. A simple example that confirms this assertion is: $\alpha=(1,0,0,0)$, $\beta=(0,1,1,0)$ and so $\alpha <_{wt} \beta$, whereas $\alpha$ and $\beta$ are incomparable with respect to the relation ''$\preceq$''. Therefore $R_{\preceq}\subset R_{<_{wt}}$. 

	We can enumerate the maximum chains in the POSet $(\{0,1\}^n, R_{\preceq})$ with the help of the next assertion.
\begin{lemma}
\label{L10}
	Let $\alpha$ be an arbitrary vector of the layer $L_{n,k}$, for some integer $k, 0< k< n$. Then $\alpha$ has $k$ adjacent vectors in the layer $L_{n,k-1}$ and also $n-k$ adjacent vectors in the layer $L_{n,k+1}$.
\begin{proof}
\normalfont{
	Let $\beta \in L_{n,k}$ be an arbitrary vector such that $\beta$ contains units in the coordinates $i_1, i_2,\dots, i_k$, where $1\leq i_1\leq \dots \leq i_k\leq n$. The set of all vectors adjacent to $\beta$ is partitioned into two subsets. The first one contains all vectors $\alpha$, such that $\alpha \preceq \beta$, i.e., exactly one of the coordinates $i_1, i_2,\dots, i_k$ is inverted to zero and all remaining coordinates are the same. Hence, there are $k$ such vectors and they are elements of $L_{n,k-1}$. The second subset contains all vectors $\gamma$, such that $\beta \preceq \gamma$, i.e., all coordinates $i_1, i_2,\dots, i_k$ are ones and exactly one of the remaining $n-k$ coordinates is inverted to one. So the number of all these vectors is $n-k$ and they belong to $L_{n,k+1}$.
}
\end{proof}
\end{lemma}
\begin{theorem}
\label{T50}
	The number of maximum chains in the POSet $(\{0,1\}^n, R_{\preceq})$ is equal to $n!$\,, for any $n\in \Nplus$.
\begin{proof}
\normalfont{
	Obviously, the length of any maximum chain is equal to the number of layers in $\{0,1\}^n$, which is $n+1$. Let $\vec{0}_n, \alpha_1, \dots, \alpha_k, \dots, \alpha_{n-1}, \vec{1}_n$ be a maximum chain. Starting from the vector $\vec{0}_n$ and following Lemma \ref{L10}, there are $n$ possible ways to choose the vector $\alpha_1\in L_{n,1}$ which is adjacent to $\vec{0}_n$. There are $n-1$ possible ways to choose a vector $\alpha_2\in L_{n,2}$ which is adjacent to $\alpha_1$, etc. There are $k$ ways to choose a vector $\alpha_k\in L_{n,k}$ which is adjacent to $\alpha_{k-1}$, etc. Finally, the last vector $\vec{1}_n$ can be chosen in a unique way. Applying the multiplication rule we obtain that the number of  maximum chains is $n.(n-1)\dots k \dots 2.1= n!$.
}
\end{proof}
\end{theorem}

	The values of $n!$, for $n=0, 1, 2, \dots$, form the sequence A000142 (called Factorial numbers) in the OEIS \cite{SLO}. Among its numerous comments, only one corresponds to the assertion of Theorem \ref{T50}. It was done on Feb 05 2006 by Rick L. Shepherd as follows: ``The number of chains of maximal length in the power set of {1, 2, ..., n} ordered by the subset relation.''. Beside the assertion of Theorem \ref{T50}, one more comment was added to the sequence A000142---it contains \textit{the number of all shortest paths} (obtained by Breadth First Search, for example) between the nodes $\vec{0}_n$ and $\vec{1}_n$ in the graph $H_n$.
\section{The weight-lexicographic order relation and two generating algorithms}
\label{WLO_Relation}
\vspace{-4pt}
	To solve the problem formulated in Section \ref {Intro} we need the serial numbers of the vectors in the sequence of layers instead of the vectors themselves. So, we shall represent the WO of $\{0,1\}^n$ by the sequence of  serial numbers of the vectors in the layers, in accordance with Theorem \ref{T10}. For that purpose, for an arbitrary layer $L_{n,k}=\{\alpha_0, \alpha_1, \dots, \alpha_m\}$ of $\{0,1\}^n$, we denote by $l_{n,k}=\#\alpha_0, \#\alpha_1, \dots, \#\alpha_m$ the \textit{sequence of serial numbers}, corresponding to the vectors of $L_{n,k}$. If $l_n= l_{n,0}, l_{n,1}, \dots, l_{n,n}$ denotes the \textit{sequence of all serial numbers}, corresponding to the vectors in the sequence of layers $L_{n,0}, L_{n,1}, \dots, L_{n,n}$, then $l_n$ represents a WO of the vectors of $\{0,1\}^n$. Briefly, we refer to $l_n$ as a \textit{WO sequence} of $\{0,1\}^n$. We note that any of all possible $\prod_{k=0}^n \binom{n}{k}!$ WO sequences can be used in solving the considered problem. But one of them deserves a special attention and we propose two algorithms for its generating. The first one (called simply \textit{Algorithm 1}) is similar to the known \textit{Bucket sort} algorithm  \cite{CLRS}. We consider each sequence $l_{n,k}$ as a bucket $B[k]$ for all vectors of weight $k$, for $k=0,1, \dots, n$. For more clarity and convenience, we assume that the buckets are represented by lists. Here is the pseudocode of Algorithm 1.
\\
\textbf{Algorithm 1.} Computing the sequence $l_n$. \\
\textbf{Input:} the integer $n\in \Nplus$. \\
\textbf{Output:} the sequence $l_n$. \\
\textbf{Procedure:} \\
1. Precomputing: following Theorem \ref{T10} and explanations after it, compute and store the weights of the vectors of $\{0,1\}^n$ in the array $wt$. So, for $i=0,1,$ $\dots, 2^n-1$, set $wt[i]= weight(\alpha)$, where $\#\alpha=i$. \\
2. Initialization: for $k=0,1, \dots, n$, set $B[k]=$ $NULL$ (i.e., empty list). Set $ln= NULL$. \\
3. Filling in the buckets: for $i=0,1,\dots, 2^n-1$, append the number $i$ to the end of $B[wt[i]]$. \\
4. Concatenation: for $k=0,1,\dots,n$, append the list $B[k]$ to the end of $ln$. \\
5. Return $ln$. \\

Notes and comments on Algorithm 1:
\begin{itemize}
	\item Its procedure and the explanations above imply its correctness.
	\item Its time complexity is a sum of the time complexities of its steps 1--4. Thus we obtain $\Theta (2^n)+ \Theta(n)+ \Theta(2^n) + \Theta(n)= \Theta(2^n)$, which is of exponential type with respect to the size of the input. It can not be better since it produces an output of exponential size. But more important is that the algorithm has a linear time complexity with respect to the size of the output. 
	\item In step 3, append the number $i$ to $B[wt[i]]$ means that if $wt[i]=k$, then the integer $i$ is appended to $B[k]$.
	\item Sorting buckets is not necessary and such a step is omitted. Any permutation into any bucket does not change WO of the entire list (sequence). 
\end{itemize}
	
	In step 3 it is written ``append $\dots$ to the end $\dots$''. Because of this and since the vectors are processed (by their serial numbers) in lexicographic order (see the cycle ``for $\dots$''), the integers in any bucket will be sorted. So the numbers in each subsequence $l_{n,k}$ are in strictly increasing order, for $0\leq k\leq n$, which means that the corresponding vectors in  $L_{n,k}$ are in lexicographic order, for $0\leq k\leq n$. Hence, after step 4 a total weight order for the sequence $l_n$ is obtained, where the lexicographic order is a second criterion for ordering the vectors of equal weights. We call it a \textit{Weight-Lexicographic Order} (WLO) and then \textit{WLO algorithm 1} is a more correct name of Algorithm 1.

	Let us continue with the mathematical bases for the second algorithm. We need the following operation on a sequence of integers.
\begin{definition}
\label{D100}
\normalfont{
	Let $n, m\in \N^+$ and $s= a_1, a_2, \dots, a_n$ be a sequence of integers. We define the operation \textit{addition of the natural number $m$ to the sequence $s$} as follows:  $s+m= a_1+m, a_2+m, \dots, a_n+m$.
}
\end{definition}

	This operation can be seen in Figure \ref{fig:Lex_N_Wt-Relation}. Following the idea in this figure and Definition \ref{D10}, we define one special WO sequence $l_n$ inductively.
\begin{definition}
\label{D110}
\normalfont{
	1) The WO sequence of the one-dimensional Boolean cube is $l_1= 0, 1$.
\\
	2) Let $l_{n-1}= l_{n-1,0}, l_{n-1,1}, \dots, l_{n-1,n-1}$ be the WO sequence of the $(n-1)$-dimensional Boolean cube.	
\\
	3) The WO sequence of $n$-dimensional Boolean cube $l_n= l_{n,0}, l_{n,1}, \dots, l_{n,n}$ is defined as follows: 
	
	$\bullet$ $l_{n,0}= 0$ and it corresponds to the layer $L_{n,0}= \{\vec{0}_n\}$; 
	
	$\bullet$	$l_{n,n}= 2^n-1$ and it corresponds to the layer $L_{n,n}= \{\vec{1}_n\}$;
	
	$\bullet$	$l_{n,k}= l_{n-1,k},\, l_{n-1,k-1}+2^{n-1}$, for $k=1, 2, \dots, n-1$. Here $l_{n,k}$ is a concatenation of two sequences: the sequence $l_{n-1,k}$ is taken (or copied) firstly, and the sequence $l_{n-1,k-1} + 2^{n-1}$ follows after it. The sequence $l_{n,k}$ corresponds to the layer $L_{n,k}$.
}
\end{definition}

	The corresponding recursive definition of $l_n$ is:
\begin{center}
\begin{tabular}{l}
   If $n=1$, then $l_1=0, 1$\,.\\
   If $n>1$, then $l_n= l_{n,0}, \dots, l_{n,k},\dots, l_{n,n}$,\ where:\\
 	 $l_{n,k}= \left\{
\begin{array}{l}
	0, \textrm{\ if\ } k=0,\\
	2^n-1, \textrm{\ if\ } k=n,\\
	l_{n-1,k},\, l_{n-1,k-1}+2^{n-1}, \textrm{\ for\ } 0 < k < n\,.\\
\end{array}
\right. 
$
\end{tabular}
\end{center}

\begin{figure}[ht]
   \centering
       \scalebox{0.55}{\includegraphics{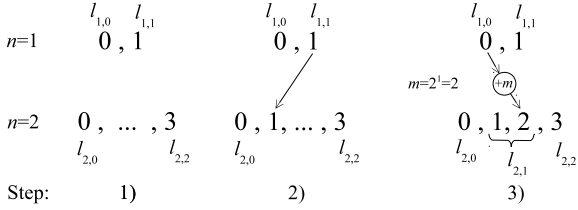}}
   \caption{The WO sequence $l_2$, obtained from $l_1$}
   \label{fig:Seq_Triangle-2}    
\end{figure}

\begin{figure}[ht]
   \centering
       \scalebox{0.53}{\includegraphics{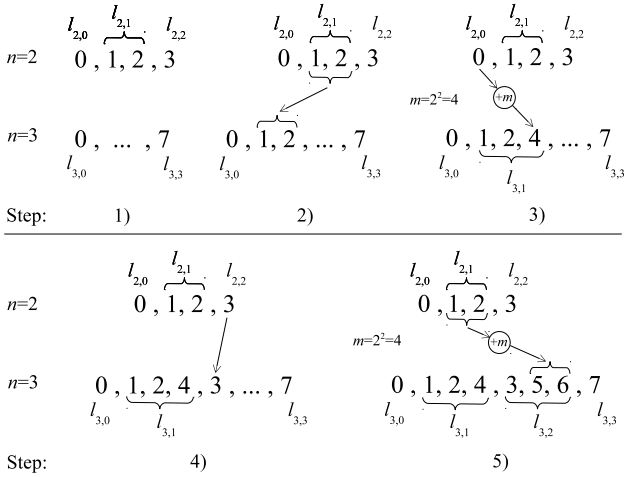}}
   \caption{The WO sequence $l_3$, obtained from $l_2$}
   \label{fig:Seq_Triangle-3}    
\end{figure}

	Figure \ref{fig:Seq_Triangle-2} and Figure \ref{fig:Seq_Triangle-3} illustrate how the WO sequences $l_2$ and $l_3$ are obtained in accordance with Definition \ref{D110}.

	The last two definitions resemble the definition of Pascal's triangle. As we noted, the length of $l_{n,k}=\binom{n}{k}=|L_{n,k}|$, for $k=0, 1, \dots, n$. Instead of the rule $\binom{n}{k}=\binom{n-1}{k-1}+\binom{n-1}{k}$ used in Pascal's triangle, we use a similar rule $l_{n,k}= l_{n-1,k},\, l_{n-1,k-1}+2^{n-1}$. The next theorem clarifies it. 
\begin{theorem}
\label{T60}
	Let $n\in \Nplus$ and $l_n= l_{n,0}, l_{n,1}, \dots, l_{n,n}$ be the WO sequence, obtained in accordance with Definition \ref{D110}. Then $l_n$ represents the vectors of $\{0,1\}^n$ in WLO.
\begin{proof}
\normalfont{
	We prove the theorem by induction on $m$, $m\in \Nplus$, following Definition \ref{D110}.

1) For $m=1$ the assertion is obvious.

2) Suppose that the theorem holds for an arbitrary integer $m>1$: the sequence $l_m= l_{m,0},l_{m,1},$ $\dots, l_{m,m}$ obtained in accordance with Definition \ref{D110} represents the vectors of $\{0,1\}^m$ in WLO.

3) Let $l_{m+1}= l_{m+1,0}, l_{m+1,1}, \dots, l_{m+1,m+1}$ be the sequence, obtained in accordance with Definition \ref{D110}. For $l_{m+1,0}=0$ and $l_{m+1,m+1}=2^{m+1}-1$, the corresponding layers $L_{m+1,0}= \{\vec{0}_{m+1}\}$ and $L_{m+1,m+1}=\{\vec{1}_{m+1}\}$ are in lexicographic order. Furthermore, $l_{m+1,0}$ and $l_{m+1,m+1}$ are in their right places in $l_{m+1}$. Let $l_{m+1,k}$ be one of the rest of the subsequences in $l_{m+1}$, for an arbitrary integer $k$, $1\leq k\leq m$. In accordance with Definition \ref{D110}, $l_{m+1,k}$ is a concatenation of two subsequences: $l_{m,k}$ and $l_{m,k-1}+2^m$, placed in that order. So, the layer $L_{m+1,k}$ corresponding to $l_{m+1,k}$ is partitioned into two groups. The first one consists of all vectors of $L_{m+1,k}$, that begin with zero. Hence their serial numbers coincide with these in the sequence $l_{m,k}$. It corresponds to the layer $L_{m,k}$, whose vectors are in lexicographic order, in accordance with the inductive suggestion. So the vectors in the first group are also in lexicographic order. The second group includes all vectors of $L_{m+1,k}$ that begin with one. So their serial numbers are obtained by an addition of the integer $2^m$ to the serial numbers of the sequence $l_{m,k-1}$. Following the inductive suggestion, the vectors of the corresponding layer $L_{m,k-1}$ are in lexicographic order and therefore the vectors in the second group are also in lexicographic order. Moreover, each vector from the first group precedes lexicographically each vector from the second group. Therefore, the sequence $l_{m+1,k}$ determines a lexicographic order into the corresponding layer $L_{m+1,k}$. This conclusion holds for any integer $k$, $1\leq k\leq m$. So, the subsequences $l_{m+1,0}, l_{m+1,1}, \dots, l_{m+1,m+1}$ represent the corresponding layers in a WO. In addition, each subsequence determines a lexicographic order into the corresponding layer. Therefore the sequence $l_{m+1}$ represents the vectors of  $\{0,1\}^{m+1}$ in a WLO. So the theorem is proven.
}
\end{proof}
\end{theorem}

	The second algorithm that we developed is called \textit{WLO algorithm 2}. It computes the sequence $l_n$ for a given input $n\in \Nplus$. The algorithm uses an array (denoted by \verb#P_t#) for the binomial coefficients from Pascal's triangle that represents the lengths of the subsequences and one more array (denoted by \verb#ss_beg#) where the beginning of each subsequence is computed and stored. The values in these two arrays are computed firstly. The code of the corresponding function is simple and it is omitted. WLO algorithm 2 is based on Definition \ref{D110}. Starting from $l_1$ it computes consecutively the sequences $l_2, l_3, \dots, l_n$ in the array \texttt{seqs}, as shown by the C/C++ programming language.
{\footnotesize
\begin{lstlisting}[language=C, caption=Computing the WLO sequence $l_n$]
typedef unsigned int uint;
void fill_in_seqs (int n) {
  seqs[1][0]= 0; // initialization
  seqs[1][1]= 1; // for n=1: l_1 
  uint m= 2; // to be added to a subseq.
  for (int r= 2; r<=n; r++) {
    seqs[r][0]= 0;
    uint k=1; 
    for (int c=1; c<=r; c++) { 
      // Preparing for step 1
      uint seq_len= P_t[r-1][c];       
      uint ssbeg= ss_beg[r-1][c]; 
      // step 1 - copying a subsequence
      for (uint j=0; j<seq_len; j++)  
        seqs[r][k++]= seqs[r-1][ssbeg+j]; 
      // Preparing for step 2
      seq_len= P_t[r-1][c-1];        
      ssbeg= ss_beg[r-1][c-1];	
      // step 2 - add m to a subsequence.
      for (uint j=0; j<seq_len; j++) 
        seqs[r][k++]= seqs[r-1][ssbeg+j]+m;
    }
    m *= 2;
  }
}
\end{lstlisting}
}

	Some results obtained by the WLO algorithms, for $n=1,2, \dots, 5$, are given in  Table \ref{tab:Results-WLO-Alg}. More results can be seen in the OEIS \cite{SLO}, sequence A294648. The results represented above were used in its creation.
\begin{table}[ht]
\footnotesize{		
	\begin{tabular}{l|l}
	$n$ & $l_n$\\
		\hline
		1 & 0, 1 \\
    2 & 0, 1, 2, 3 \\
    3 & 0, 1, 2, 4, 3, 5, 6, 7 \\
    4 & 0, 1, 2, 4, 8, 3, 5, 6, 9, 10, 12, 7, 11, 13, 14, 15 \\
    5 & 0, 1, 2, 4, 8, 16, 3, 5, 6, 9, 10, 12, 17, 18, 20, 24, 7, $\dots$ \\ 
    % 11,13, 14, 19, 21, 22, 25, 26, 28, 15, 
		\hline
		\end{tabular}
}		
	\caption{Results obtained by the WLO algorithms, for $n=1,2,\dots, 5$}
	\label{tab:Results-WLO-Alg}
\end{table}
	
	As we said, WLO algorithm 2 is based on Definition \ref{D110} and follows its steps. This definition and Theorem \ref{T60} determine its correctness. Let us compute the time complexity of the algorithm. The time for filling in both additional arrays (\verb#P_t#, for Pascal's triangle and \verb#ss_beg#, for the beginning of each subsequence) is proportional to the number of integers that they contain, i.e., $\Theta (n^2)$. The function \texttt{fill\_in\_seqs} in Listing 1 runs as follows. On the $k$-th step, $2\leq k\leq n$, it copies generally $2^{k-1}-1$ values from $l_{k-1}$ to $l_k$, and also it adds the constant $m=2^{k-1}$ to $2^{k-1}-1$ members of $l_{k-1}$ and stores them in $l_k$. So, it performs $\Theta(2^k)$ assignments and $\Theta(2^{k-1})$ summations, i.e., $\Theta(2^k)$ operations generally on the $k$-th step. Therefore, the time complexity of the algorithm is
\[ 
\sum_{k=2}^n\Theta(2^k)= \Theta \left( \sum_{k=2}^n 2^k\right)= \Theta(2^{n+1})=  \Theta(2^n)\,.
\]

	So, the time complexity of WLO algorithm 2 is of the same type as at WLO algorithm 1. Let us consider the space complexity of WLO algorithm 2. For clarity, in Listing 1 we use a two-dimensional array of size $2^n\times 2^n$ and hence, the space complexity is $\Theta(2^{2n})$. We recall that the existence of $l_k$ is sufficient to obtain $l_{k+1}$. So, instead of the square array we can use:
\begin{itemize}
	\item Two one-dimensional static arrays of size $2^n$---for the existing sequence $l_k$ and for the new sequence $l_{k+1}$. After we obtain $l_{k+1}$, we change the role of the arrays to obtain the next sequence $l_{k+2}$, and so on.
	\item One-dimensional arrays of size $2^k$ which are created/deleted dynamically in the $k$-th step, for $k=1,2, \dots, n$.
\end{itemize}
In both cases the space complexity of the WLO algorithm reduces to $\Theta(2^n)$. 

		Let us return to the VectorOfMaxWeight problem and comment on the usage of WLO sequence $l_n$ in its solving. Let $f$ be a Boolean function of $n$ variables given by its true table vector $TT(f)= (y_0,y_1,\dots, y_{2^n-1})$. We want to find (if exists) a vector $\alpha \in \{0,1\}^n$ of maximal weight, such that $f(\alpha)=1$. After we know enough about the WLO, we do not comment more on the trivial approach (exhaustive search). The efficient solving of this problem consists of consecutive checks of the coordinates of $TT(f)$ in accordance with the WLO sequence $l_n$, from the last to the first term of it. Let $k$ be the first number of a non-zero coordinate in $TT(f)$. Then $k=\#\alpha$ such that $f(\alpha)=1$, $\alpha$ has a maximal weight and all other such vectors (if exist) preceded lexicographically $\alpha$. This is seen in the following C/C++ code, where the array $f$ stands for $TT(f)$, and the arrays $wt$ and $ln$ mean the same as in WLO algorithm 1.
{\footnotesize
\begin{lstlisting}[language=C, caption=Search by the WLO sequence $l_n$]
int vector_of_max_weight (bool f[size]) {
  for (int i= size - 1; i>=0; i--) {	
    int k= ln[i]
    if (f[k])
      return k; // and wt[k] if necessary
  }
  return -1; // when f is the 0-constant
}
\end{lstlisting}
}	 

\hyphenation{Vec-tor-Of-Max-Weight}	 
	The distribution of Boolean functions according to their algebraic degrees given in \cite{VB2019} and \cite[sequence A319511]{SLO} shows that when $n$ grows, almost $100\%$ of all Boolean functions of $n$ variables have algebraic degree $n$ or $n-1$ (i.e., exists a vector of weight $n$ or $n-1$  which is a solution to the VectorOfMaxWeight problem). So, this search will finish after no more than $n+1$ checks at almost $100\%$ of all such functions. But the general time complexity of this search is proportional to the length of WLO sequence $l_n$, and so it is $O(2^n)$. For brevity, we call this algorithm \textit{Search by WLO}.
	 
\begin{figure}[ht]
   \centering
       \scalebox{0.58}{\includegraphics{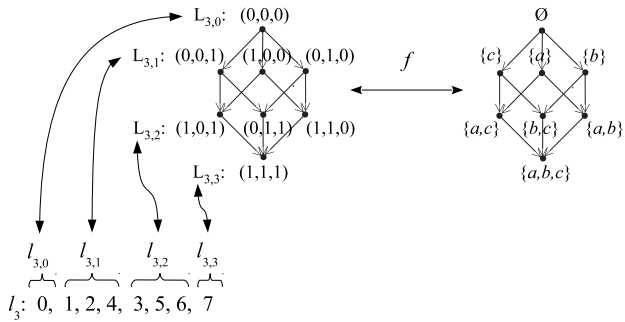}}
   \caption{Illustration of the bijections between the sequence $l_3$, the layers of $\{0,1\}^3$ and the subsets of $\{a, b, c\}$, as well as the isomorphism between the POSets $(\{0,1\}^3,\preceq)$ and $({\mathcal{P}}(\{a,b,c\}), \subseteq)$}
   \label{fig:WLO-B_cube-Subsets}    
\end{figure}

	Finally, we note that the bijection between the $n$-dimensional Boolean cube and the power set of a given $n$-element set (Theorem \ref{T20}) means that both WLO algorithms can have more general applications. For example, they can be used in solving problems related to representing and generating the subsets of a given set in a certain order, or some of its subsets (for example, all subsets of $k$ elements, or $k$-combinations), etc., as it is shown in Figure \ref{fig:WLO-B_cube-Subsets}. In such cases, the WLO sequence considered as a sequence of serial numbers of characteristic vectors means a cardinality order of the subsets. Furthermore, if the elements of the set are in lexicographic order, the corresponding subsets of equal size will be in reverse lexicographic order. Figure \ref{fig:WLO-B_cube-Subsets} summarizes some of the discussed results and illustrates:
\begin{itemize}
	\item the bijection between subsequences of $l_3$ and the layers of $\{0,1\}^3$;
	\item the bijection $f$ between the vectors of $\{0,1\}^3$ and the subsets of $\{a,b,c\}$ (see Theorem \ref{T20});
	\item the isomorphism $f$ between the POSets $(\{0,1\}^3, \preceq)$ and $({\mathcal{P}}(\{a,b,c\})$, $\subseteq )$ by the graphs of the corresponding relations.
\end{itemize}
\section{Characteristic vectors (masks) of the layers and their generating}
\label{Gen_CV_WO_Rel}
\vspace{-4pt}
	So far we have considered the representation of the vectors from the layer $L_{n,k}$ by the corresponding sequence $l_{n,k}$ of their serial numbers, for $k=0,1,$ $\dots, n$. Another way for representation is by the \textit{characteristic vector} $m_{n,k}$ \textit{of the layer} $L_{n,k}$, for $k=0,$ $1,\dots, n$ (the denotation $m$ comes from ``mask'' and will be understood later). Following  Definition \ref{D130}, we define it as follows: $m_{n,k}= (a_0,a_1, \dots, a_{2^n-1})\in \{0,1\}^{2^n}$, where:
\begin{eqnarray*}
a_i=\left\{
\begin{array} {ll}
0, \textrm{\ if\ } \alpha_i\notin L_{n,k}\,,\\
1, \textrm{\ if\ } \alpha_i\in L_{n,k}\,,
\end{array}
\right.
\end{eqnarray*} 	
$\alpha_i\in \{0,1\}^n$, for $i=0, 1, \dots ,2^n-1$.	
	
	Since the characteristic vectors are binary vectors, they can be represented in a bitwise manner in one or several compute words. For example, for given $n, d\in \Nplus$, where $d$ is the number of bits in one computer word, there are $n+1$ characteristic vectors, and each of them will occupy $s$ computer words, where $s=1$ if $n\leq d$, or $s=2^{n-d}$ if $n>d$. 
	
	Let $f$ be a Boolean function of $n$ variables given by its $TT(f)$. This vector can be represented in a byte-wise way or in a bitwise way. The byte-wise representation of $TT(f)$ is used in the Search by WLO algorithm (see Listing 2). So it is natural to think about its bitwise version of this algorithm. It is possible to check the bits of $TT(f)$ in accordance with the WLO sequence $l_n$. Then the corresponding algorithm will be similar to the byte-wise Search by WLO algorithm and it will have the same time complexity. However, we can check all vectors in the same layer in one (or several, say $s$) step(s). For this purpose, we shall use the characteristic vectors, since $m_{n,k}$ is a binary vector of the same length as $TT(f)$ and $m_{n,k}$ contains units only in these bits that correspond to the integers in the subsequence $l_{n,k}$ (and hence these units correspond to the vectors in $L_{n,k}$), for $k=0,1,\dots,n$. So, $m_{n,k}$ masks only the significant bits for $L_{n,k}$, $0\leq k\leq n$, and that is why the characteristic vectors are called \textit{masks} in this application. Thus, we need to repeat bitwise conjunctions between $TT(f)$ and $m_{n,i}$, for $i=n, n-1,\dots, 0$, until the result of the serial conjunctions is zero. The first index $k, 0\leq k \leq n$, for which $TT(f)\wedge m_{n,k}>0$ means that there are one or more vectors in $L_{n,k}$, such that $f$ takes a value $1$ on each of them. If the algorithm returns $k$ (the maximal weight) and stops, this is enough for computing the algebraic degree of a Boolean function. But this is a solution to the restricted version of the VectorOfMaxWeight problem since the algorithm does not return a vector. When we need a vector, we have to process the result of $TT(f)\wedge m_{n,k}$ when it becomes $>0$ for the first time. 
	
	Let us comment on some details of the version that computes only the maximal weight. We call it \textit{Bitwise Search by WLO} algorithm, accepting that it always uses masks. When $TT(f)$ occupies one computer word, the algorithm performs at most $n+1$ steps and so its time complexity is $O(n)$, i.e., it is of logarithmic type ($n=\log_2{2^n}$) with respect to the size of the input. If the size of computer word is $64=2^6$ bits and $f$ is a function of $n>6$ variables, then $TT(f)$ occupies $s=2^{n-6}$ computer words. So, $m_{n,i}$ will occupy $s$ computer words also and the computing of $TT(f) \wedge m_{n,i}$ will be done in $s$ steps, for $i=n, n-1, \dots, 0$. If on some of these steps the conjunction between the corresponding computer words of $TT(f)$ and $m_{n,i}$ is greater than zero, the algorithm returns $i$ and stops. Therefore, in this (general) case, the time complexity of the algorithm becomes $O(n).O(s)=O(n.2^{n-6})$. This is seen in the following C/C++ code, where the masks are represented by a two-dimensional array. The number of its rows is equal to the number of variables ($n\_vars$) $+ 1$, and the number of columns is equal to the number of computer words ($n\_cwords$) used for the representation of $TT(f)$ (the variable $f$ in the code).
{\footnotesize
\begin{lstlisting}[language=C, caption=Bitwise Search by WLO]
typedef unsigned long long ull;
int max_deg_by_masks (ull f[]) {
  for (int row= n_vars; row >= 0; row--) {
    for (int col= 0; col<n_cwords; col++) {	
      if (f[col] & masks[row][col])			
        return row; // the layer's number
    }
  }
  return -1; // when f is the 0-constant
}
\end{lstlisting}
}

	It is important to consider the masks' generating. For arbitrary $i, 0\leq i \leq n$, it is easy to put units in all these bits of $m_{n,i}$ that correspond to the numbers in the subsequence $l_{n,i}$. So this way of generating the masks has a time complexity which is proportional to the length of WLO sequence $l_n$, i.e., $\Theta(2^n)$. 
	
	We propose one more way to generate the masks. As at the layers, we use the serial numbers of the masks $\#m_{n,i}$ instead of their vectors $m_{n,i}$, for $0\leq i \leq n$. They will be obtained and stored in the necessary number of 64-bits computer words---as many as for the $TT(f)$ vector. So, we can generate them in accordance with the following definition.
\begin{definition}
\label{D30}
\normalfont{
1) For $n=1$, the serial numbers of the masks corresponding to the subsequences $l_{1,0}$ and $l_{1,1}$ are $\#m_{1,0}=2$ and $\#m_{1,1}=1$.
\\
2) Let $\#m_{n-1,0}, \#m_{n-1,1}, \dots, \#m_{n-1,n-1}$ be the serial numbers of the masks corresponding to the subsequences $l_{n-1,0}, l_{n-1,1}, \dots, l_{n-1,n-1}$. 
\\
3) The serial number of the mask $\#m_{n,i}$ corresponding to the subsequence $l_{n,i}$ is:
\begin{eqnarray*}
\#m_{n,i}=\left\{
\begin{array} {lr}
2^{2^{n-1}}.\#m_{n-1,0}=2^{2^n-1}, \textrm{\ if\ } i=0\,, \\
1, \textrm{\ if\ } i=n\,, \\
2^{2^{n-1}}.\#m_{n-1,i} + \#m_{n-1,i-1}, \\ 
\hskip 38mm
\textrm{\ if\ } 0<i<n\,,
\end{array}
\right.
\end{eqnarray*}
for $i=0,1, \dots ,n$.
}
\end{definition}
	
	Definition \ref{D30} corresponds to definitions \ref{D10} and \ref{D110}. Its correctness can be proven strictly by mathematical induction. The algorithm for masks' generating based on this definition, as well as the previous one, have some particularities when $n>6$ and they work with $s=2^{n-6}$ computer words. The running time for generating (precomputing) the masks by each of these two algorithms is  negligible ($\approx 0$ seconds). The serial numbers of the masks grow exponentially, as it is seen in Table \ref{Ta2:Masks4vars}.
\begin{table}[ht]
  \centering
  \small{
  \begin{tabular}{|c|c|c|c|c|c|}
  \hline
	$n=$ & $\#m_{n,0}$ &  $\#m_{n,1}$ &  $\#m_{n,2}$ &  $\#m_{n,3}$ &  $\#m_{n,4}$ \\
	\hline
	1 & 2 & 1 & -- & -- & -- \\
	\hline
	2 & 8 & 6 & 1 & -- & -- \\
	\hline
	3 & 128 & 104 & 22 & 1 & -- \\
	\hline
	4 & 32768 & 26752 & 5736 & 278 & 1 \\
	\hline
	\end{tabular}
	\caption{The serial numbers of the masks, for $n=1,\dots ,4$}
  \label{Ta2:Masks4vars}
}	
\end{table}	

  These and some additional results were used in creating the sequence A305860 in OEIS \cite{SLO}. 
\begin{example}
\label{Ex20}
\normalfont{
	We shall illustrate how the problem considered is solved by the Byte-wise and Bitwise Search by WLO algorithms, for a Boolean function $f$ of $4$ variables. Its $TT(f)$, the coordinates' numbers and the masks (for $n=4$) are given in Table \ref{Ta3:Example}. When we use the Byte-wise WLO Algorithm, it checks consecutively the coordinates of $TT(f)$, from right to left, i.e., 15, 14, 13, 11, 7, 12---see the WLO sequence $l_4$ in Table \ref{tab:Results-WLO-Alg}. $TT(f)f$ contains zeros in all coordinates before 12-th, but $TT(f)$ contains one in this coordinate and so the algorithm stops after 6 checks. Since 12 is a term of $l_{4,2}$, hence the vector of maximal weight ($=2$) has a serial number $12$. When the Bitwise WLO Algorithm is used, it computes the conjunctions: $TT(f)\wedge m_{4,4}=0$, $TT(f)\wedge m_{4,3}=0$, $TT(f)\wedge m_{4,2}>0$ and thereafter it stops. So a vector of weight 2 is a solution to the problem and it is computed in 3 steps. As a continuation outside the algorithm, the vector $TT(f)\wedge m_{4,2}= (0, 0, 0, 1, 0, 1, 1, 0, 0, 0, 1, 0, 1, 0, 0, 0)$ contains units in coordinates $3, 5, 6, 10, 12$. They are the serial numbers of all vectors from $\{0,1\}^4$ that have weight $2$ and $f$ takes a value 1 on each of them.
}
\begin{table}[ht]
  \centering
  \footnotesize{
  \begin{tabular}{|r|r|r|}
  \hline
	             Coordinates' numbers & 0 1 2 3 4 5 6 7 & 8 9 0 1 2 3 4 5 \\
	\hline
	                         $TT(f)=$ & 1 0 0 1 0 1 1 0 & 1 0 1 0 1 0 0 0 \\
	\hline
	$\#m_{4,0}=32768,\ m_{4,0}=$      & 1 0 0 0 0 0 0 0 & 0 0 0 0 0 0 0 0 \\
	\hline
	$\#m_{4,1}=26752,\ m_{4,1}=$      & 0 1 1 0 1 0 0 0 & 1 0 0 0 0 0 0 0 \\
	\hline
	$\#m_{4,2}=5736,\hfill  m_{4,2}=$ & 0 0 0 1 0 1 1 0 & 0 1 1 0 1 0 0 0 \\
	\hline
	$\#m_{4,3}=278,\hfill m_{4,3}=$   & 0 0 0 0 0 0 0 1 & 0 0 0 1 0 1 1 0 \\
	\hline
	$\#m_{4,4}=1,\hfill m_{4,4}=$     & 0 0 0 0 0 0 0 0 & 0 0 0 0 0 0 0 1 \\
	\hline
	\end{tabular}
  \caption{The data used in Example \ref{Ex20}}
  \label{Ta3:Example}
}	
\end{table}		
\end{example}
\section{Experimental results}
\label{Exp_Res}
\vspace{-4pt}
	We developed three algorithms for solving a version of the VectorOfMaxWeight problem where the algorithms compute the maximal weight of a vector $\alpha$ such that $f(\alpha)=1$. Their names Exhaustive search, Byte-wise Search by WLO and Bitwise Search by WLO show how they work. We conducted a series of tests to compare the efficiency of these algorithms after we know their theoretical time complexities, i.e., to understand what is the difference between $\Theta(2^n)$ and $O(2^n)$ at this problem. All tests were executed on the same computer (Intel Pentium CPU G4400, 3.3 GHz, 4GB RAM, Samsung SSD 650 120 GB), at the same conditions. The programs were developed (and executed) under Windows 10 OS and MVS Express 2015 for Windows Desktop. They are written in C++ programming language, built in Release mode as 32-bits and 64-bits console applications and executed without an Internet connection. All tests were executed 3 times and the running times are taken on average. The $TT(f)$ vectors of all tested Boolean functions are represented in a byte-wise and bitwise manner. In the next tables, the time for  conversion between these types of representation, as well as the time for reading from a file, are excluded from the running times. The time for generating (precomputing) the WLO sequence $l_n$ and the masks is negligible and it also is excluded.

	Table \ref{tab:Comparison5vars} shows the pure running time of the algorithms, for all $2^{32}$ Boolean functions of 5 variables. 
\begin{table}[ht]
  \centering
  \small{
  \begin{tabular}{|c|c|c|c|}
  \hline
	              & \multicolumn{3}{c |}{Pure running time in seconds for:} \\
		            \cline{2-4} 
  Implemen-     & Exhaustive & Byte-wise & Bitwise   \\
   tation       &	Search     & S. by WLO & S. by WLO \\		            
  \hline
  32 bits appl. & 130.292    & 39.989    & 3,053 \\
  \hline
  64 bits appl. & 174.819    & 45.786    & 3,240 \\
  \hline
	\end{tabular}
	}
	\caption{Experimental results for all $2^{32}$ Boolean functions of 5 variables} 
  \label{tab:Comparison5vars}
\end{table}

	To test Boolean functions of 6 and more variables, we created and used a file of $10^8$ randomly generated 64-bits unsigned integers (the file size is $\approx 1.39$ GBytes). If $n>6$ then $2^{n-6}$ integers are read from the file and thus they form the consecutive Boolean function. Tables \ref{tab:More_vars32} and \ref{tab:More_vars64} show the pure running time of the algorithms being compared, for Boolean functions (BFs) of 6 and more variables.
\begin{table}[!ht]
  \centering
  \small{
  \begin{tabular}{|c|c|c|c|}
  \hline
Number of & \multicolumn{3}{c |}{Pure running time in seconds for:} \\
		  \cline{2-4} 
variables & Exhaustive & Byte-wise & Bitwise   \\
and BFs   &	Search    & S. by WLO & S. by WLO   \\
  \hline
6 vars,   & 18.371 & 1.540 & 0,507 \\
$10^8$ BFs & & & \\
  \hline
8 vars,   & 22.758 & 0.380 & 0,703 \\
$10^8/4$ BFs & & & \\
  \hline
10 vars,  & 24.130 & 0.224 & 0,177 \\
$10^8/16$ BFs & & & \\
  \hline    
12 vars,  & 25.310 & 0.074 & 0.074 \\
$10^8/64$ BFs & & & \\
  \hline      
16 vars,  & 26.000 & 0.068 & 0.070 \\
$97656$ BFs & & & \\
  \hline      
	\end{tabular}
}
	\caption{Experimental results for the 32 bits implementations}
	\label{tab:More_vars32}
\end{table}
\begin{table}[!ht]
  \centering
  \small{
  \begin{tabular}{|c|c|c|c|}
  \hline
Number of & \multicolumn{3}{c |}{Pure running time in seconds for:} \\
		  \cline{2-4} 
variables & Exhaustive & Byte-wise & Bitwise   \\
and BFs   &	Search    & S. by WLO & S. by WLO   \\
  \hline
6 vars,   & 21.802 & 1.816 & 0,583 \\
$10^8$ BFs & & & \\
  \hline
8 vars,   & 23.381 & 0.083 & 0,045 \\
$10^8/4$ BFs & & & \\
  \hline
10 vars,  & 23.520 & 0.045 & 0,181 \\
$10^8/16$ BFs & & & \\
  \hline    
12 vars,  & 24.574 & 0.029 & 0,125 \\
$10^8/64$ BFs & & & \\
  \hline      
16 vars,  & 25.911 & 0.015 & 0.203 \\
$97656$ BFs & & & \\
  \hline      
	\end{tabular}
}
	\caption{Experimental results for the 64 bits implementations}
	\label{tab:More_vars64}
\end{table}

	These results clearly show the benefits of the WLO approaches in solving the VectorOfMaxWeight problem. The reader can make his own conclusions about the efficiency depending on the algorithm chosen, its implementation, the number of variables, etc. 
\section{Conclusions}
\label{Concl}
\vspace{-4pt}%
	We should note, that the problem considered has very specific searching space---almost $100\%$ of all Boolean functions of $n$ variables are of degree $n$ or $n-1$, as it is shown in \cite{VB2019} and \cite[sequence A319511]{SLO}. This fact explains the efficiency of algorithms based on WLO, as well as why the Byte-wise Search by WLO becomes faster than the Bitwise one when $n$ grows. As it was shown, in almost $100\%$ of all cases the first algorithm finishes after at most $n+1$ steps, whereas the second---after at most $n+1$ steps, when $n\leq 6$, or at most $2^{n-6}+1$ steps when $n>6$. Thus $(n+1)$ becomes smaller than $2^{n-6}+1$ when $n>9$ and this explains the results in tables \ref{tab:More_vars32} and \ref{tab:More_vars64} obtained by the algorithms for searching by WLO.

	We note that if the Boolean function is represented by the vector of coefficients of its algebraic normal form instead of its $TT(f)$, then the algorithms will compute the algebraic degree of this Boolean function. We have already shown that the computing of this important cryptographic parameter can be performed more efficiently by using WLO based algorithms.
	
	At the end of Section \ref{WLO_Relation} we discussed some other applications of the algorithms for generating the WLO sequence---in representing and generating all subsets of a given set in a definite order, or some of its subsets (for example, all $k$-element subsets, or $k$-combinations), etc. The same applications can have the algorithms for generating the masks considered in Section \ref{Gen_CV_WO_Rel}. It is important to note that the bijection between the subsets, their characteristic vectors and their serial numbers is a very simple and convenient ranking/unranking function. All these algorithms use and process only the serial numbers in generating, i.e., they do not generate the objects but their ranked representation. Furthermore, they do this very efficiently.
\vspace{2pt}
\\
\textbf{Acknowledgments:}	The author is grateful for the partial support from the Research Fund of the University of Veliko Tarnovo, Bulgaria, under Contract FSD-31-340-14/26.03.2019.

\end{document}